\documentstyle[11pt]{article} 
 
\begin{document}

\begin{titlepage} 
\begin{flushright} 
{OUTP-98-66-P}\\ 
{submitted to Phys. Lett. B}\\ 
\end{flushright} 
\vskip 0.5 cm 
\begin{center} 
 {\Large{\bf Unification and Extra Space-time Dimensions}}\\ 
        \vskip 0.6 cm 
        {\large{\bf Dumitru Ghilencea\footnote{ 
                E-mail address: D.Ghilencea1@physics.oxford.ac.uk} 
               } and {\bf 
                Graham G. Ross\footnote{ 
                E-mail: G.Ross1@physics.oxford.ac.uk}}}\\ 
        \vskip 1 cm 
{{\it Department of Physics, Theoretical Physics, University of Oxford}}\\ 
{\it 1 Keble Road, Oxford OX1 3NP, United Kingdom}\\ 
\vskip 0.2 cm 
 
\end{center} 
\vskip 2 cm 
\vskip 2 cm 
\begin{abstract} 
We analyse the phenomenological implications of a particular class of 
supersymmetric models with additional  
space-time dimensions below the unification scale.  
Assuming the unification 
of the gauge couplings and using a two-loop 
calculation below the scale of the additional 
space-time dimensions, we show that the   
value of $\alpha_3(M_z)$ is further increased from the two-loop 
Minimal Supersymmetric Standard  Model prediction. We consider whether  
decompactification threshold effects  
could bring $\alpha_3(M_z)$ into agreement with experiment 
and discuss the associated  
level of fine tuning needed. 
\end{abstract} 
\end{titlepage} 
 
\setcounter{footnote}{0} 
 
\section{Introduction} 
 
There has been much recent interest \cite{dienes_v1,dienes_v2,dienes_v3} in 
the possibility that the unification of the gauge couplings of the Standard 
Model could take place at a scale significantly below that of the Minimal 
Supersymmetric Standard Model (MSSM). In the MSSM, with four dimensional 
gauge coupling running, the energy scale at which the gauge couplings unify 
is given by $M_{g}^o\approx 1-3\times 10^{16}$ GeV. However it is possible 
that unification could occur at a scale much lower than that of the MSSM 
through the presence  of extra space-time dimensions 
\cite{dienes_v1,dienes_v2,dienes_v3,marco}
of a relatively large radius $R$ associated with an 
energy scale less than $M_{g}^o$. Extra space-time dimensions appear 
naturally in string theory, and therefore, such an idea has a strong 
mathematical motivation. The way this idea is implemented in practice is by 
introducing towers of Kaluza-Klein excitations associated with the gauge 
bosons, the Higgs fields and, possibly, with the generations of fermions. 
The exact details of how this is done and the difficulties encountered in 
this are to some extent model dependent. Here we consider the case of ref.  
\cite{dienes_v1,dienes_v2,dienes_v3} where such a model was presented in 
detail. We will restrict ourselves to the case that preserves supersymmetry 
as a main ingredient, although there are scenarios where the 
non-supersymmetric Standard Model is valid up to the scale of the extra 
space-time dimensions. 
 
The purpose of this work is to analyse the implications of such models for 
the value of the unification scale and the strong coupling at the 
electroweak scale. The latter prediction provides one of the few 
quantitative tests of unification. While it is remarkably successful the 
MSSM prediction is now significantly above the upper experimental value. It 
is particularly interesting therefore to determine whether models with extra 
space-time dimensions can improve on this. 
 
\section{Results from the Renormalisation Group Evolution} 
 
In the following we present a two-loop renormalisation group calculation of 
the running of the gauge couplings to make predictions for the strong 
coupling at electroweak scale and the value $\mu _{0}$ of the mass scale at 
which the new extra space-time dimensions appear. The value of the 
unification scale $\Lambda $ is also computed. We also take into account the 
effect of low-energy supersymmetric thresholds on our predictions. 
 
\subsection{Previous results} 
 
First we present the equations derived in ref.~\cite{dienes_v1} which 
describe the evolution of the couplings above the decompactification scale $%
\mu _{0}$. The equations were derived via an effective theory approach with $%
\Lambda $ as the ultraviolet cut-off. It denotes the scale at which physics 
(e.g. string excitations) beyond the higher dimensional theory must be 
included. The resulting evolution is given by 
\cite{dienes_v1,dienes_v2,dienes_v3}  
\begin{equation} 
\alpha _{i}^{-1}(\mu _{0})=\alpha _{i}^{-1}(\Lambda )+\frac{b_{i}-{\tilde{b}%
_{i}}}{2\pi }\ln \frac{\Lambda }{\mu _{0}}+\frac{\tilde{b}_{i}}{4\pi }I(\mu 
_{0},\Lambda ,\delta )  \label{gen1} 
\end{equation} 
Equation (\ref{gen1}) gives the value of the gauge couplings at $\mu _{0}$ 
for any $\Lambda \geq \mu _{0}$. In this equation, $b_{i}=(33/5,1,-3)$ are 
the usual MSSM beta functions. Also  
\begin{equation} 
\tilde{b}_{i}=(3/5,-3,-6)+\eta (4,4,4) 
\end{equation} 
The coefficients $\tilde{b}_{i}$ correspond to the contributions of the 
appropriate Kaluza-Klein states at each massive Kaluza-Klein excitations 
level. The effect of $\eta $ families of matter in complete $SU(5)$ 
vector-like representations is accounted for in $\tilde{b}_{i}$ by the term 
proportional to $\eta$. In the following, unless explicitly stated, we 
assume that $\eta \not=0$. We also have \cite{dienes_v1,dienes_v2,dienes_v3}  
\begin{equation} 
I(\mu _{0},\Lambda ,\delta )=\int_{r\Lambda ^{-2}}^{r\mu _{0}^{-2}}\frac{dt}{%
t}\left[ \vartheta _{3}\left( 0,e^{{-t}/{R^{2}}}\right) \right] ^{\delta } 
\label{gen2} 
\end{equation} 
where \cite{erdelyi}  
\begin{equation} 
\vartheta _{3}\left( u,q\right) =\sum_{n=-\infty }^{n=\infty 
}q^{n^{2}}e^{i2nu} 
\end{equation} 
is the elliptic Jacobi theta function and \cite 
{dienes_v1,dienes_v2,dienes_v3}  
\begin{equation} 
r=\left[ \Gamma (1+\delta /2)\right] ^{2/\delta } 
\end{equation} 
Here $\delta $ represents the number of extra dimensions considered, $\delta 
=D-4$, while $R$ represents the radius of the extra space-time dimensions 
which is identified \cite{dienes_v1,dienes_v2,dienes_v3} with the inverse of 
the mass scale $\mu _{0}$, $R\equiv 1/\mu _{0}$. The integral of eq.(\ref 
{gen2}) can also be written as  
\begin{equation} 
I(\mu _{0},\Lambda ,\delta )\equiv {\cal J}(\Lambda /\mu _{0},\delta 
)=\int_{r/(\Lambda /\mu _{0})^{2}}^{r}\frac{dx}{x}\left[ \vartheta 
_{3}\left( 0,e^{-x}\right) \right] ^{\delta }  \label{integral} 
\end{equation} 
which therefore depends on the ratio $\Lambda /\mu _{0}$ only. 
 
A useful approximation for the above integral is $\vartheta 
_{3}(0,e^{-x})\approx \sqrt{\frac{\pi }{x}}$ which gives  
\begin{equation} 
{\cal J}(\Lambda /\mu _{0},\delta )\approx \frac{2X_{\delta }}{\delta }\left[ 
\left( \frac{\Lambda }{\mu _{0}}\right) ^{\delta }-1\right]  \label{approxim} 
\end{equation} 
with $X_{\delta }$ standing for the following quantity  
\begin{equation} 
X_{\delta }=\frac{\pi ^{\delta /2}}{\Gamma (1+\delta /2)} 
\end{equation} 
The difference between the two quantities given in eq.(\ref{integral}) and 
eq.(\ref{approxim}) respectively is small for $\delta =1,2,3$ and can be 
safely ignored (for $\delta =1,2,3$ it gives an error in $\alpha _{3}(M_{z})$ 
less than $10^{-5}$ ). For larger $\delta $, the approximation of the Jacobi 
function introduces a larger error for $\alpha _{3}(M_{z})$ for the case of 
a large ratio $\Lambda /\mu _{0}$. 
 
To avoid this we will use in our calculation below the 
exact form for ${\cal J}$, 
the integral of Jacobi function, eq.(\ref{integral}). With the above 
observations, eq.(\ref{gen1}) becomes \cite{dienes_v1,dienes_v2,dienes_v3}  
\begin{equation} 
\alpha _{i}^{-1}(\mu _{0})=\alpha _{i}^{-1}(\Lambda )+\frac{b_{i}-{\tilde{b}%
_{i}}}{2\pi }\ln \frac{\Lambda }{\mu _{0}}+\frac{\tilde{b}_{i}}{2\pi }\frac{1%
}{2}{\cal J}\left(\frac{\Lambda }{\mu _{0}},\delta\right)  \label{RGEmu0} 
\end{equation} 
 
\subsection{Two-loop RGE results} 
 
We proceed now to derive the implications of the above model for the value 
of the strong coupling at electroweak scale and for the unification scale. 
We use the values of $\alpha _{i}(\mu _{0})$ of eq.(\ref{RGEmu0}) as input 
for the equations which describe the running of the gauge couplings below 
the scale $\mu _{0}$ where a MSSM spectrum applies\footnote{%
By the MSSM spectrum we mean the quarks and leptons sector with their 
superpartners, the gauge plus gaugino sector and two $SU(2)$ Higgs doublets 
(no Higgs triplets) and their superpartners.}. Below the scale $\mu _{0}$ we 
employ the two-loop RGE evolution for the gauge couplings  
\begin{equation} 
\alpha _{i}^{-1}(M_{z})=-\Delta _{i}+\alpha _{i}^{-1}(\mu _{0})+\frac{b_{i}}{%
2\pi }\ln \left[ \frac{\mu _{0}}{M_{Z}}\right] +\frac{1}{4\pi }\sum_{j=1}^{3}%
\frac{b_{ij}}{b_{j}}\ln \left[ \frac{\alpha _{j}(\mu _{0})}{\alpha 
_{j}(M_{Z})}\right]  \label{RGE} 
\end{equation} 
This is the integral form for the running of the gauge couplings below $\mu 
_{0}$ scale, valid in two-loop order, with a MSSM-like spectrum and standard 
model gauge symmetry. This can be easily derived by integrating the two-loop 
differential equations for the running of the gauge couplings with a 
MSSM-like spectrum below the scale $\mu _{0}$ and with upper values for the 
gauge couplings equal to $\alpha _{j}(\mu _{0})$. The coefficients $b_{i}$ 
and $b_{ij}$ denote the one-loop and two-loop beta functions which are just 
those of the MSSM; $\Delta _{i}$ include the low-energy supersymmetric 
thresholds and ${\overline{MS}}\rightarrow {\overline{DR}}$ scheme 
conversion factors\footnote{%
The scheme conversion factors will cancel in the calculation below as we 
compare our results to those of the MSSM model. The same applies to the low 
energy supersymmetric thresholds.}. From eqs.(\ref{RGEmu0}),(\ref{RGE}) we 
obtain  
\begin{equation} 
\alpha _{i}^{-1}(M_{z})=-\Delta _{i}+\alpha _{\Lambda }^{-1}+\frac{b_{i}}{%
2\pi }\ln \left[ \frac{\Lambda }{M_{Z}}\right] +\frac{\tilde{b}_{i}}{2\pi }%
\left\{ \frac{1}{2}{\cal J}\left( \frac{\Lambda }{\mu _{0}},\delta \right) 
-\ln \frac{\Lambda }{\mu _{0}}\right\} +\frac{1}{4\pi }\sum_{j=1}^{3}\frac{%
b_{ij}}{b_{j}}\ln \left[ \frac{\alpha _{j}(\mu _{0})}{\alpha _{j}(M_{Z})}%
\right]  \label{fullRGE} 
\end{equation} 
where we have now {\it imposed} the unification of the gauge couplings at 
the scale $\Lambda $. 
 
To predict $\alpha_{3}(M_{Z})$ we input the well measured values for $%
\alpha _{1}(M_{Z})$ and $\alpha _{2}(M_{Z}).$ However, to make detailed 
predictions we need some information about the low energy supersymmetric 
thresholds, $\Delta _{i}$. These have been discussed in the context of the 
MSSM  \cite{lan}. Since these effects are associated with the
low-energy structure of the theory, they remain the same in the models
considered here. Thus we choose to eliminate them by relating 
the new predictions to those of the MSSM.
 
\bigskip In the MSSM model (the MSSM variables are labelled with an ``o'' 
index to distinguish them from the model based on extra space-time 
dimensions)  
\begin{equation} 
\alpha _{i}^{o-1}(M_{z})=-\Delta _{i}+\alpha _{g}^{o-1}+\frac{b_{i}}{2\pi }%
\ln \left[ \frac{M_{g}^{o}}{M_{Z}}\right] +\frac{1}{4\pi }\sum_{j=1}^{3}%
\frac{b_{ij}}{b_{j}}\ln \left[ \frac{\alpha _{g}^{o}}{\alpha _{j}^{o}(M_{Z})}%
\right]  \label{MSSM} 
\end{equation} 
with $\alpha_g^o\equiv \alpha_j^o(M_g^o)$.
We substitute the $\Delta _{i}$ of eq.(\ref{MSSM}) into eq.(\ref{fullRGE}%
). This gives the three equations presented below, where as mentioned, we 
take $\alpha _{1}(M_{z})=\alpha _{1}^{o}(M_{z})$ and $\alpha 
_{2}(M_{z})=\alpha _{2}^{o}(M_{z})$ from experiment. We have (i=1,2)  
\begin{eqnarray} \label{al12}
0 &=&\alpha _{\Lambda }^{-1}-\alpha _{g}^{o-1}+\frac{b_{i}}{2\pi }\ln \frac{ 
\Lambda }{M_{g}^{o}}+\frac{\tilde{b}_{i}}{2\pi }\left\{ \frac{1}{2}{\cal J}
\left( \frac{\Lambda }{\mu _{0}},\delta \right) -\ln \frac{\Lambda }{\mu _{0} 
}\right\}  \nonumber \\ 
&+&\frac{1}{4\pi }\sum_{j=1}^{3}\frac{b_{ij}}{b_{j}}\ln \left[ \frac{\alpha 
_{j}(\mu _{0})}{\alpha _{g}^{o}}\right] +\frac{1}{4\pi }\frac{b_{i3}}{b_{3}} 
\ln \left[ \frac{\alpha _{3}^{o}(M_{z})}{\alpha _{3}(M_{z})}\right] \\ 
&&  \nonumber \\ 
\alpha _{3}^{-1}(M_{z})-\alpha _{3}^{o-1}(M_{z}) &=&\alpha _{\Lambda 
}^{-1}-\alpha _{g}^{o-1}+\frac{b_{3}}{2\pi }\ln \frac{\Lambda }{M_{g}^{o}}+ 
\frac{\tilde{b}_{3}}{2\pi }\left\{ \frac{1}{2}{\cal J}\left(
\frac{\Lambda }
{\mu _{0}},\delta \right) -\ln \frac{\Lambda }{\mu _{0}}\right\}  \nonumber \\ 
&+&\frac{1}{4\pi }\sum_{j=1}^{3}\frac{b_{3j}}{b_{j}}\ln \left[ \frac{\alpha 
_{j}(\mu _{0})}{\alpha _{g}^{o}}\right] +\frac{1}{4\pi }\frac{b_{33}}{b_{3}} 
\ln \left[ \frac{\alpha _{3}^{o}(M_{z})}{\alpha _{3}(M_{z})}\right] 
\label{al3} 
\end{eqnarray} 
Equations (\ref{al12}) and (\ref{al3}) represent a system of three equations 
where the values for $\alpha _{j}(\mu _{0})$ are taken from eq.(\ref{RGEmu0}%
) with $\alpha _{i}(\Lambda )\equiv \alpha _{\Lambda }$. We consider as 
input parameters the values of $\eta $, $\delta $ and the ratio $\rho \equiv 
\ln (\Lambda /\mu _{0})$. The output of these equations is represented
by $\alpha_{3}(M_{z})$, $\mu_{0}$ and $\alpha_{\Lambda }$. 
 
The full numerical results following from these equations are given in 
Tables 1,2,3,4. We may see that $\alpha _{3}(M_{Z})$ is systematically 
increased compared to the MSSM value, while the unification scale is 
decreased. We may understand the structure of these results analytically. 
From eqs.({\ref{al12}), subtracting the case $i=1$ and $i=2$, gives  
\begin{equation} 
\ln \frac{\Lambda }{M_{g}^{o}}=-\frac{9}{14}\left\{ \frac{1}{2}{\cal J}%
\left( \frac{\Lambda }{\mu _{0}},\delta \right) -\ln \frac{\Lambda }{\mu _{0}%
}\right\} +Two\,\,loop\,\,contribution 
\end{equation} 
Imposing that $\Lambda >M_{z}$ gives $\Lambda /M_{g}^{o}>(1/3)\times 
10^{-14} $ (with $M_{g}^{o}\approx 3\times 10^{16}GeV$), 
and we find the following (one-loop) {\it upper limits} to the ratio $%
\Lambda /\mu _{0}$ (independent of the value of $\eta $)  
\begin{eqnarray} 
\frac{\Lambda }{\mu _{0}} &\approx &28.5\,\,\,\,\,if\,\,\,\,\delta =1 \\ 
\frac{\Lambda }{\mu _{0}} &\approx &5.9\,\,\,\,\,if\,\,\,\,\delta =2 \\ 
\frac{\Lambda }{\mu _{0}} &\approx &3.4\,\,\,\,\,if\,\,\,\,\delta =3 
\end{eqnarray} 
This means that the energy range between $\mu _{0}$ and $\Lambda $ is 
relatively small. This is due to the steep behaviour in the evolution of the 
couplings, introduced by the power-law contribution (of function ${\cal J}$, 
see eq.(\ref{approxim})), which also has the effect of {\it increasing} $%
\alpha _{3}(M_{z})$ from the MSSM value as we will discuss later. } 
 
From eqs.({\ref{al12}) for $i=1$ and $i=2$ we also get that  
\begin{equation} 
\alpha _{\Lambda }^{-1}=\alpha _{g}^{o-1}+\frac{51-56\eta }{28\pi }\left\{  
\frac{1}{2}{\cal J}\left( \frac{\Lambda }{\mu _{0}},\delta \right) -\ln  
\frac{\Lambda }{\mu _{0}}\right\} +Two\,\,loop\,\,contribution 
\end{equation} 
}or, using the approximation (\ref{approxim}),  
\begin{equation} 
\alpha _{\Lambda }^{-1}=\alpha _{g}^{o-1}+\frac{51-56\eta }{28\pi }\left\{  
\frac{X_{\delta }}{\delta }\left[ \left( \frac{\Lambda }{\mu _{0}}\right) 
^{\delta }-1\right] -\ln \frac{\Lambda }{\mu _{0}}\right\} 
+Two\,\,loop\,\,contribution 
\end{equation} 
{This means that in the absence of the extra-matter ($\eta =0$) the value of  
$\alpha _{\Lambda }$ decreases as we increase (for fixed $\delta $) the 
ratio $\Lambda /\mu _{0}$. In the presence of the extra-matter ($\eta
\not=0$) the value of $\alpha _{\Lambda }$ increases with 
the ratio $\Lambda /\mu_{0}$.  
Two-loop terms can affect this observation, but not significantly, as may be 
seen from the full two loop results of Tables 1,2,3,4. } 
 
To analyse the implications for $\alpha _{3}(M_{z})$ we add together eq.(%
\ref{al3}) and eq.(\ref{al12}) for $i=1$ multiplied by $%
(b_{2}-b_{3})/(b_{1}-b_{2})$ and with eq.(\ref{al12}) for $i=2$ multiplied 
by $(b_{3}-b_{1})/(b_{1}-b_{2})$. This has the effect of eliminating the 
contributions from $\alpha _{\Lambda }^{-1}$ and $\ln \Lambda $ terms to the 
difference $\alpha _{3}^{-1}(M_{z})-\alpha _{3}^{o-1}(M_{z})$. The result 
obtained in this way is 
\begin{eqnarray}
\alpha_3^{-1}(M_z)-\alpha_3^{o -1}(M_z)&=&-\frac{3}{14\pi}\left\{
\frac{1}{2} {\cal J}\left(\frac{\Lambda}{\mu_0},\delta\right)
-\ln \frac{\Lambda}{\mu_0}\right\}+\sum_{j=1}^{3}\omega_j\ln\alpha_j(\mu_0)
\nonumber\\
&&
+\frac{477}{70\pi}\ln\alpha_g^o
+\frac{17}{14\pi}\ln\frac{\alpha_3^o(M_z)}{\alpha_3(M_z)}\label{twoloop}
\end{eqnarray}
with $\omega_j$ given by
\begin{equation}
\omega_j=\left\{\frac{2}{11\pi}, -\frac{15}{2\pi}, \frac{17}{14\pi}\right\}_j
\end{equation}
with $\alpha _{3}^{o}(M_{z})\approx 0.126$ and $\alpha _{g}^{o}\approx 
0.0433 $. 
 
For large values of $\mu _{0}$, very close to the $\Lambda $ scale, the two 
loop terms in the model with extra dimensions are close to those of the MSSM 
and they cancel in eq.(\ref{twoloop}), while the explicit (power-law) 
term in the same equation has a less important role. In this case the lhs of 
eq.(\ref{twoloop}) is close to 0. Thus, in the limit $\mu _{0}=\Lambda  
$ we get $\alpha _{3}(M_{z})=0.126$ as in the MSSM. This can also be seen in 
the (full two-loop) results presented in Tables~1,2,3,4, obtained from 
solving numerically eqs.(\ref{al12}) and (\ref{al3}). As the ratio $\Lambda 
/\mu _{0}$ increases, the power-law correction of ${\cal J}(\Lambda /\mu 
_{0},\delta )$ in eq.(\ref{twoloop}) dominates, with the effect of 
increasing $\alpha _{3}(M_{z})$. This effect can again be seen from the 
numerical results of Tables~1,2,3,4, for various input parameters $\eta $, $%
\delta $ and $\rho \equiv \ln (\Lambda /\mu _{0})$. 
 
For the case of lower values of $\mu _{0}$ we may get a first indication of 
the result by ignoring the last three subdominant (two-loop) terms
in eq.(\ref{twoloop}). Then the 
right hand side of the above equation gives a negative contribution for $%
\Lambda \geq \mu _{o}$. (Note the behaviour of ${\cal J}$ with $\Lambda /\mu 
_{0}$, given in eq.(\ref{approxim})). Therefore, the value of the strong 
coupling at electroweak scale in the model with extra space-time dimensions 
is increased above its corresponding MSSM value. As we ignored the two-loop 
terms in the rhs of eq.(\ref{twoloop}), the lhs of this equation 
stands for a one-loop approximation, and thus, $\alpha 
_{3}(M_{z})|_{one-loop}>\alpha _{3}^{o}(M_{z})|_{one-loop}\approx 0.117$. In 
general, two-loop contributions tend to increase the prediction for the 
strong coupling from its one-loop value. To be more explicit consider eq.(%
\ref{twoloop}) again. The dominant contribution in the above equation is the 
curly bracket, and it increases $\alpha _{3}(M_{z})$ as discussed. The term $%
\ln \alpha _{g}^{o}$ is also negative and it increases $\alpha _{3}(M_{z})$. 
The term $17/(14\pi )\ln (\alpha _{3}^{o}(M_{z})/\alpha _{3}(M_{z}))$ could 
lower the value of the strong coupling but its contribution is very small.  
The only contribution which has a lowering effect on the strong coupling 
comes from the term with $j=2$ under the sum over $j$, as $\ln \alpha 
_{j}(\mu _{0})<0$ and is insufficient to reduce $\alpha _{3}(M_{z})$ below 
the MSSM value. Note that $\alpha _{j}(\mu _{0})$ can be further replaced by 
its expression given in eq.(\ref{RGEmu0}), and therefore the expression of 
the strong coupling given above depends on $\alpha _{\Lambda }$ and on the  
{\it ratio} $\Lambda /\mu _{0}>1$ only. 
 
\bigskip 
 
\bigskip 
 
What happens if we add extra-matter in complete $SU(5)$
representations\footnote{
The effect of extra matter in complete $SU(5)$ representations is to increase
$\alpha_3(M_z)$ \cite{grl} due to two-loop terms. However, in the present
model, above the scale $\mu_0$ where extra matter in vector-like
representations is assumed to exist, there are no two loop 
contributions from extra-matter states because of the underlying
$N=2$ symmetry.}? 
The answer is easily obtained from eq.(\ref{RGEmu0}) which we write below 
in the following 
form (with $\alpha _{i}(\Lambda )\equiv \alpha _{\Lambda }$)  
\begin{equation} 
\alpha _{i}^{-1}(\mu _{0})=\alpha _{\Lambda }^{-1}+\frac{4\eta }{2\pi }%
\left\{ \frac{1}{2}{\cal J}\left( \frac{\Lambda }{\mu _{0}},\delta \right) 
-\ln \frac{\Lambda }{\mu _{0}}\right\} +\frac{b_{i}-{\tilde{b}_{i}}(\eta =0)%
}{2\pi }\ln \frac{\Lambda }{\mu _{0}}
+\frac{\tilde{b}_{i}(\eta =0)}{2\pi}
\frac{1}{2}{\cal J}\left(\frac{\Lambda}{\mu_0},\delta\right)
\label{alpj} 
\end{equation} 
The value of $\alpha _{i}(\mu _{0})$ in eq.(\ref{alpj}) remains unchanged 
in the presence of the extra-matter if we rescale $\alpha _{\Lambda }$ by  
\begin{equation} 
\alpha _{\Lambda }^{-1}=\alpha _{\Lambda }^{-1}(\eta =0)-\frac{4\eta }{2\pi }%
\left\{ \frac{1}{2}{\cal J}\left( \frac{\Lambda }{\mu _{0}},\delta \right) 
-\ln \frac{\Lambda }{\mu _{0}}\right\}  \label{rescale} 
\end{equation} 
or, using the approximation of eq.(\ref{approxim})  
\begin{equation} 
\alpha _{\Lambda }^{-1}=\alpha _{\Lambda }^{-1}(\eta =0)-\frac{4\eta }{2\pi }%
\left\{ \frac{X_{\delta }}{\delta }\left[ \left( \frac{\Lambda }{\mu _{0}}%
\right) ^{\delta }-1\right] -\ln \frac{\Lambda }{\mu _{0}}\right\} 
\label{rescale2} 
\end{equation} 
If we apply the rescaling in eqs.(\ref{al12}) and (\ref{al3}), we obtain 
three equations similar in form to those in the absence of the extra-matter, 
but with a new $\alpha _{\Lambda }$. The conclusion is that our numerical 
predictions for $\alpha _{3}(M_{z})$, $\mu _{0}$ and $\Lambda $ for the case  
$\eta =0$ will remain the same when $\eta {\not=}0$, as can also be seen by 
comparing the results of Tables~1,2,3,4 for various values for $\eta $. The 
only effect is a change (increase) in the value of $\alpha _{\Lambda }$ by 
an amount given in equation (\ref{rescale}). For the strong coupling, this 
effect can also be seen in eq.(\ref{twoloop}), where only $\alpha _{j}(\mu 
_{0})$ depends on the presence of the extra-matter, and this dependence is 
re-absorbed into a redefinition of $\alpha _{\Lambda }$. 
 
The predictions we made in the presence of the extra-matter are valid as 
long as the extra-matter decouples at the scale $\mu _{0}$.  
However, extra matter does not necessarily decouple at the scale $\mu _{0}$ 
since these states, being vector-like under the Standard Model gauge group, 
are not protected by any chiral symmetry and are therefore very heavy, of 
mass $M\geq \mu _{0}$. Introducing a new parameter $M$ for the mass of these 
vector-like states would make our analysis less tractable, and for this 
reason we restricted ourselves to the case when $M=\mu _{0}$ and various 
values for $\eta $. 
 
Given the discrepancy between the predictions and experimental
measurements\footnote{The experimental value \cite{particledata}
 for $\alpha_3(M_z)$ is $0.118\pm 0.003$.} for $\alpha_3(M_z)$ 
one may ask whether there are effects that could reconcile the two. The 
value of the strong coupling at the electroweak scale is very sensitive to 
the thresholds for the various Kaluza Klein excitations. One may ask whether 
such threshold effects could be the origin of the discrepancy
\cite{dienes_v3}. Thus we now estimate how such threshold effects can 
accommodate a change of the strong coupling (at the $M_{z}$ scale) large enough 
to bring it close to the experimental value, within the experimental 
accuracy of $\pm 0.003$, with a unification scale in the region of few TeV. 
To do this, we allow for different threshold effects in $\alpha_i$,
$i=\{1,2,3\}$, by introducing in eq.(\ref{fullRGE}) different threshold
scales $\mu_{0;i}$ with $i=\{1,2,3\}$. It is sufficient for our purposes
to consider $\mu_{0;3}$ only, so that eq.(\ref{al12}) remains unchanged.
 The leading change in the 
strong coupling at the electroweak scale is then given by  
\begin{equation} 
\Delta \alpha _{3}^{-1}(M_{z})\approx \frac{{\tilde{b}_{3}}}{2\pi }\frac{%
X_{\delta }}{\delta }\left[ \frac{\Lambda }{\mu _{0; 3}}\right] ^{\delta }%
\frac{\Delta \mu _{0; 3}}{\mu _{0; 3}}  \label{bra} 
\end{equation} 
We note (see Table~1) that for $\mu_{0}$ in the region of few TeV the 
strong coupling is given by $\alpha _{3}(M_{z})\approx 0.17$ (for
$\delta =1$), so a change of 0.05 is needed to 
bring it to the experimentally measured 
value. This corresponds to $\Delta \alpha _{3}^{-1}(M_{z})\approx 3.47$, 
giving 
\begin{equation} 
\frac{\Delta \mu _{0; 3}}{\mu _{0; 3}}\approx \frac{1}{14} 
\end{equation} 
The accuracy, ${\left[ \frac{\Delta \mu_{0; 3}}{\mu_{0;3}}\right] _{w},}$ to 
which this must hold to keep within the experimental error of $0.003$ 
requires $\Delta \alpha _{3 w}^{-1}(M_{z})\approx 0.21$. From eq.(\ref{bra}) 
we find  
\begin{equation} 
{\left[ \frac{\Delta \mu _{0;3}}{\mu _{0;3}}\right] _{w}}\left[ \frac{\Delta \mu 
_{0;3}}{\mu _{0;3}}\right] ^{-1}\approx \frac{\Delta \alpha _{3w}^{-1}(M_{z})}{%
\Delta \alpha _{3}^{-1}(M_{z})}\approx \frac{0.21}{3.47}\approx \frac{1}{16} 
\end{equation} 
Thus  
\begin{equation} 
\frac{\Delta \mu _{0;3}}{\mu _{0;3}}\approx \frac{1}{14}\left( 1\pm \frac{1}{16}%
\right) \approx \frac{1}{14}\pm \frac{1}{224} 
\end{equation} 
This means that in order to keep the strong coupling within $0.003$ of $0.12$ 
for a unification scale in the region of few TeV, one must fine tune the 
thresholds to one part in 224. 
 
\begin{table}[tbp] 
\begin{center} 
\begin{tabular}{|l|l|c|c|c|l|c|} 
\hline 
$\eta$ & $\delta$ & $\rho=\ln(\Lambda/\mu_0)$ & $\Lambda$ (GeV) & $\mu_0$ 
(GeV) & $\alpha_3(M_Z)$ & $\alpha_\Lambda$ \\ \hline\hline 
0 & 1 & 0. & $3\times 10^{16}$ & $3\times 10^{16}$ & 0.1260 & 0.0433 \\  
\hline 
0 & 1 & 0.2 & $2.55\times 10^{16}$ & $2.09\times 10^{16}$ & 0.1261 & 0.0430 
\\ \hline 
0 & 1 & 0.4 & $2.03\times 10^{16}$ & $1.36\times 10^{16}$ & 0.1263 & 0.0426 
\\ \hline 
0 & 1 & 0.6 & $1.51\times 10^{16}$ & $8.26\times 10^{15}$ & 0.1266 & 0.0421 
\\ \hline 
0 & 1 & 0.8 & $1.01\times 10^{16}$ & $4.54\times 10^{15}$ & 0.1271 & 0.0414 
\\ \hline 
0 & 1 & 1.0 & $6.04\times 10^{15}$ & $2.22\times 10^{15}$ & 0.1277 & 0.0407 
\\ \hline 
0 & 1 & 1.2 & $3.13\times 10^{15}$ & $9.44\times 10^{14}$ & 0.1285 & 0.0397 
\\ \hline 
0 & 1 & 1.4 & $1.37\times 10^{15}$ & $3.37\times 10^{14}$ & 0.1295 & 0.0385 
\\ \hline 
0 & 1 & 1.6 & $4.81\times 10^{14}$ & $9.71\times 10^{13}$ & 0.1309 & 0.0371 
\\ \hline 
0 & 1 & 1.8 & $1.31\times 10^{14}$ & $2.16\times 10^{13}$ & 0.1327 & 0.0356 
\\ \hline 
0 & 1 & 2.0 & $2.59\times 10^{13}$ & $3.51\times 10^{12}$ & 0.1349 & 0.0338 
\\ \hline 
0 & 1 & 2.2 & $3.48\times 10^{12}$ & $3.86\times 10^{11}$ & 0.1378 & 0.0318 
\\ \hline 
0 & 1 & 2.4 & $2.92\times 10^{11}$ & $2.65\times 10^{10}$ & 0.1416 & 0.0296 
\\ \hline 
0 & 1 & 2.6 & $1.37\times 10^{10}$ & $1.02\times 10^{9}$ & 0.1464 & 0.0273 
\\ \hline 
0 & 1 & 2.8 & $3.15\times 10^{8}$ & $1.91\times 10^7$ & 0.1528 & 0.0250 \\  
\hline 
0 & 1 & 3.0 & $3.04\times 10^{6}$ & $1.52\times 10^5$ & 0.1610 & 0.0225 \\  
\hline 
0 & 1 & 3.2 & $1.00\times 10^{4}$ & $0.41\times 10^3$ & 0.1712 & 0.0201 \\  
\hline\hline 
0 & 2 & 0. & $3\times 10^{16}$ & $3\times 10^{16}$ & 0.126 & 0.0433 \\ \hline 
0 & 2 & 0.2 & $2.06\times 10^{16}$ & $1.69\times 10^{16}$ & 0.1264 & 0.0427 
\\ \hline 
0 & 2 & 0.4 & $1.10\times 10^{16}$ & $7.40\times 10^{15}$ & 0.1271 & 0.0416 
\\ \hline 
0 & 2 & 0.6 & $4.09\times 10^{15}$ & $2.24\times 10^{15}$ & 0.1284 & 0.0401 
\\ \hline 
0 & 2 & 0.8 & $8.74\times 10^{14}$ & $3.92\times 10^{14}$ & 0.1304 & 0.0379 
\\ \hline 
0 & 2 & 1.0 & $8.20\times 10^{13}$ & $3.01\times 10^{13}$ & 0.1337 & 0.0350 
\\ \hline 
0 & 2 & 1.2 & $2.25\times 10^{12}$ & $6.79\times 10^{11}$ & 0.1391 & 0.0314 
\\ \hline 
0 & 2 & 1.4 & $9.87\times 10^{9}$ & $2.43\times 10^9$ & 0.1478 & 0.0271 \\  
\hline 
0 & 2 & 1.6 & $2.76\times 10^{6}$ & $5.58\times 10^5$ & 0.1626 & 0.0225 \\  
\hline\hline 
0 & 3 & 0. & $3\times 10^{16}$ & $3\times 10^{16}$ & 0.1260 & 0.0433 \\  
\hline 
0 & 3 & 0.2 & $1.61\times 10^{16}$ & $1.32\times 10^{16}$ & 0.1267 & 0.0423 
\\ \hline 
0 & 3 & 0.4 & $4.69\times 10^{15}$ & $3.15\times 10^{15}$ & 0.1283 & 0.0403 
\\ \hline 
0 & 3 & 0.6 & $4.45\times 10^{14}$ & $2.44\times 10^{14}$ & 0.1315 & 0.0371 
\\ \hline 
0 & 3 & 0.8 & $5.47\times 10^{12}$ & $2.46\times 10^{12}$ & 0.1379 & 0.0322 
\\ \hline 
0 & 3 & 1.0 & $1.61\times 10^{9}$ & $5.92\times 10^{8}$ & 0.1513 & 0.0260 \\  
\hline 
0 & 3 & 1.2 & $4.94\times 10^{2}$ & $1.49\times 10^{2}$ & 0.1794 & 0.0191 \\  
\hline\hline 
\end{tabular} 
\end{center} 
\caption{The (2-loop RG) results for the strong coupling at the electroweak 
scale for $\protect\delta =1,2,3$, with $\protect\eta =0.$ The parameter $%
\protect\rho $ is constrained to give $\protect\mu _{0}$ above the 
electroweak scale and below $\Lambda $. The above results remain valid if we 
change $\protect\eta $ to non-zero values, with the only difference that $%
\protect\alpha _{\Lambda }$ changes according to equation
(\ref{rescale}) bringing (for 
fixed $\protect\delta $) the unified coupling within non-perturbative region 
for a value of the $\protect\rho $ parameter as given in Table~3 and 
Table~4. The results presented are obtained for 
$\protect\alpha_{3}^{o}(M_{z})=0.126$, $\alpha_g^o=0.0433$ 
and $M_{g}^{o}=3\times 10^{16}$ GeV. } 
\label{table:1} 
\end{table} 
 
\begin{table}[tbp] 
\begin{center} 
\begin{tabular}{|l|l|c|c|c|l|c|} 
\hline 
$\eta$ & $\delta$ & $\rho=\ln(\Lambda/\mu_0)$ & $\Lambda$ (GeV) & $\mu_0$ 
(GeV) & $\alpha_3(M_Z)$ & $\alpha_\Lambda$ \\ \hline\hline 
0 & 4 & 0. & $3\times 10^{16}$ & $3\times 10^{16}$ & 0.1260 & 0.0433 \\  
\hline 
0 & 4 & 0.1 & $2.15\times 10^{16}$ & $1.95\times 10^{16}$ & 0.1264 & 0.0427 
\\ \hline 
0 & 4 & 0.2 & $1.27\times 10^{16}$ & $1.04\times 10^{16}$ & 0.1271 & 0.0419 
\\ \hline 
0 & 4 & 0.3 & $5.62\times 10^{15}$ & $4.16\times 10^{15}$ & 0.1281 & 0.0406 
\\ \hline 
0 & 4 & 0.4 & $1.61\times 10^{15}$ & $1.08\times 10^{15}$ & 0.1297 & 0.0388 
\\ \hline 
0 & 4 & 0.5 & $2.42\times 10^{14}$ & $1.47\times 10^{14}$ & 0.1324 & 0.0363 
\\ \hline 
0 & 4 & 0.6 & $1.39\times 10^{13}$ & $7.63\times 10^{12}$ & 0.1366 & 0.0331 
\\ \hline 
0 & 4 & 0.7 & $1.89\times 10^{11}$ & $9.38\times 10^{10}$ & 0.1433 & 0.0293 
\\ \hline 
0 & 4 & 0.8 & $2.98\times 10^{8}$ & $1.39\times 10^{8}$ & 0.1545 & 0.0250 \\  
\hline 
0 & 4 & 0.9 & $1.85\times 10^{4}$ & $7.51\times 10^{3}$ & 0.1731 & 0.0204 \\  
\hline\hline 
0 & 5 & 0. & $3\times 10^{16}$ & $3\times 10^{16}$ & 0.1260 & 0.0433 \\  
\hline 
0 & 5 & 0.1 & $2.04\times 10^{16}$ & $1.85\times 10^{16}$ & 0.1265 & 0.0426 
\\ \hline 
0 & 5 & 0.2 & $1.04\times 10^{16}$ & $8.53\times 10^{15}$ & 0.1273 & 0.0415 
\\ \hline 
0 & 5 & 0.3 & $3.31\times 10^{15}$ & $2.45\times 10^{15}$ & 0.1288 & 0.0398 
\\ \hline 
0 & 5 & 0.4 & $4.80\times 10^{14}$ & $3.22\times 10^{14}$ & 0.1315 & 0.0371 
\\ \hline 
0 & 5 & 0.5 & $1.91\times 10^{13}$ & $1.16\times 10^{13}$ & 0.1362 & 0.0335 
\\ \hline 
0 & 5 & 0.6 & $8.95\times 10^{10}$ & $4.91\times 10^{10}$ & 0.1446 & 0.0287 
\\ \hline 
0 & 5 & 0.7 & $1.22\times 10^{7}$ & $6.07\times 10^{6}$ & 0.1606 & 0.0232 \\  
\hline\hline 
0 & 6 & 0. & $3\times 10^{16}$ & $3\times 10^{16}$ & 0.1260 & 0.0433 \\  
\hline 
0 & 6 & 0.1 & $1.98\times 10^{16}$ & $1.79\times 10^{16}$ & 0.1265 & 0.0426 
\\ \hline 
0 & 6 & 0.2 & $9.09\times 10^{15}$ & $7.44\times 10^{15}$ & 0.1275 & 0.0413 
\\ \hline 
0 & 6 & 0.3 & $2.09\times 10^{15}$ & $1.55\times 10^{15}$ & 0.1295 & 0.0391 
\\ \hline 
0 & 6 & 0.4 & $1.37\times 10^{14}$ & $9.15\times 10^{13}$ & 0.1333 & 0.0356 
\\ \hline 
0 & 6 & 0.5 & $8.96\times 10^{11}$ & $5.44\times 10^{11}$ & 0.1409 & 0.0306 
\\ \hline 
0 & 6 & 0.6 & $8.86\times 10^{7}$ & $4.86\times 10^{7}$ & 0.1569 & 0.0242 \\  
\hline\hline 
\end{tabular} 
\end{center} 
\caption{As for Table~1 with $\protect\delta =4,5,6$.} 
\label{table:2} 
\end{table} 
 
\begin{table}[tbp] 
\begin{center} 
\begin{tabular}{|l|l|c|c|c|l|c|} 
\hline 
$\eta$ & $\delta$ & $\rho=\ln(\Lambda/\mu_0)$ & $\Lambda$ (GeV) & $\mu_0$ 
(GeV) & $\alpha_3(M_Z)$ & $\alpha_\Lambda$ \\ \hline\hline 
3 & 1 & 0. & $3\times 10^{16}$ & $3\times 10^{16}$ & 0.1260 & 0.0433 \\  
\hline 
3 & 1 & 0.2 & $2.55\times 10^{16}$ & $2.09\times 10^{16}$ & 0.1261 & 0.0439 
\\ \hline 
3 & 1 & 0.4 & $2.03\times 10^{16}$ & $1.36\times 10^{16}$ & 0.1263 & 0.0448 
\\ \hline 
3 & 1 & 0.6 & $1.51\times 10^{16}$ & $8.26\times 10^{15}$ & 0.1266 & 0.0460 
\\ \hline 
3 & 1 & 0.8 & $1.01\times 10^{16}$ & $4.54\times 10^{15}$ & 0.1271 & 0.0477 
\\ \hline 
3 & 1 & 1.0 & $6.04\times 10^{15}$ & $2.22\times 10^{15}$ & 0.1277 & 0.0502 
\\ \hline 
3 & 1 & 1.2 & $3.13\times 10^{15}$ & $9.44\times 10^{14}$ & 0.1285 & 0.0537 
\\ \hline 
3 & 1 & 1.4 & $1.37\times 10^{15}$ & $3.37\times 10^{14}$ & 0.1295 & 0.0590 
\\ \hline 
3 & 1 & 1.6 & $4.81\times 10^{14}$ & $9.71\times 10^{13}$ & 0.1309 & 0.0672 
\\ \hline 
3 & 1 & 1.8 & $1.31\times 10^{14}$ & $2.16\times 10^{13}$ & 0.1327 & 0.0815 
\\ \hline 
3 & 1 & 2.0 & $2.59\times 10^{13}$ & $3.51\times 10^{12}$ & 0.1349 & 0.1107 
\\ \hline 
3 & 1 & 2.2 & $3.48\times 10^{12}$ & $3.86\times 10^{11}$ & 0.1378 & 0.1995 
\\ \hline\hline 
3 & 2 & 0. & $3\times 10^{16}$ & $3\times 10^{16}$ & 0.126 & 0.0433 \\ \hline 
3 & 2 & 0.2 & $2.06\times 10^{16}$ & $1.69\times 10^{16}$ & 0.1264 & 0.0447 
\\ \hline 
3 & 2 & 0.4 & $1.10\times 10^{16}$ & $7.40\times 10^{15}$ & 0.1271 & 0.0473 
\\ \hline 
3 & 2 & 0.6 & $4.09\times 10^{15}$ & $2.24\times 10^{15}$ & 0.1284 & 0.0523 
\\ \hline 
3 & 2 & 0.8 & $8.74\times 10^{14}$ & $3.92\times 10^{14}$ & 0.1304 & 0.0624 
\\ \hline 
3 & 2 & 1.0 & $8.20\times 10^{13}$ & $3.01\times 10^{13}$ & 0.1337 & 0.0887 
\\ \hline 
3 & 2 & 1.2 & $2.25\times 10^{12}$ & $6.79\times 10^{11}$ & 0.1391 & 0.2460 
\\ \hline\hline 
3 & 3 & 0. & $3\times 10^{16}$ & $3\times 10^{16}$ & 0.1260 & 0.0433 \\  
\hline 
3 & 3 & 0.2 & $1.61\times 10^{16}$ & $1.32\times 10^{16}$ & 0.1267 & 0.0458 
\\ \hline 
3 & 3 & 0.4 & $4.69\times 10^{15}$ & $3.15\times 10^{15}$ & 0.1283 & 0.0516 
\\ \hline 
3 & 3 & 0.6 & $4.45\times 10^{14}$ & $2.44\times 10^{14}$ & 0.1315 & 0.0682 
\\ \hline 
3 & 3 & 0.8 & $5.47\times 10^{12}$ & $2.46\times 10^{12}$ & 0.1379 & 0.1719 
\\ \hline 
\end{tabular} 
\end{center} 
\caption{As for Table 1 for $\protect\delta=1,2,3$ and $\protect\eta=3$. For  
$\protect\rho $ outside the range presented in the Table, the coupling $%
\protect\alpha _{\Lambda }$ becomes larger than unity. Note that we must 
also have that $(\Lambda /\protect\mu _{0})^{\protect\delta }\times \protect%
\alpha _{\Lambda }\leq {\cal O}(4\protect\pi ) $ so that perturbation theory 
works well. As expected, the results for $\Lambda $, $\protect\mu _{0}$ and $%
\protect\alpha _{3}(M_{z})$ are identical to the corresponding cases of 
Table~1, the only difference being a renormalisation of the coupling 
$\protect\alpha _{\Lambda }$, as explained in eq.(\ref{rescale}) in the text. 
The results presented are obtained from eqs.(\ref{al12}) and (\ref{al3}) 
with $\protect\alpha _{g}^{o}=0.0433$, $\protect\alpha _{3}^{o}(M_{z})=0.126$ 
and $M_{g}^{o}=3\times 10^{16}$ GeV, using the integral of Jacobi function, 
eq.(\ref{integral}).} 
\end{table} 
 
\begin{table}[tbp] 
\begin{center} 
\begin{tabular}{|l|l|c|c|c|l|c|} 
\hline 
$\eta$ & $\delta$ & $\rho=\ln(\Lambda/\mu_0)$ & $\Lambda$ (GeV) & $\mu_0$ 
(GeV) & $\alpha_3(M_Z)$ & $\alpha_\Lambda$ \\ \hline\hline 
3 & 4 & 0. & $3\times 10^{16}$ & $3\times 10^{16}$ & 0.1260 & 0.0433 \\  
\hline 
3 & 4 & 0.1 & $2.15\times 10^{16}$ & $1.95\times 10^{16}$ & 0.1264 & 0.0446 
\\ \hline 
3 & 4 & 0.2 & $1.27\times 10^{16}$ & $1.04\times 10^{16}$ & 0.1271 & 0.0467 
\\ \hline 
3 & 4 & 0.3 & $5.62\times 10^{15}$ & $4.16\times 10^{15}$ & 0.1281 & 0.0507 
\\ \hline 
3 & 4 & 0.4 & $1.61\times 10^{15}$ & $1.08\times 10^{15}$ & 0.1297 & 0.0580 
\\ \hline 
3 & 4 & 0.5 & $2.42\times 10^{14}$ & $1.47\times 10^{14}$ & 0.1324 & 0.0745 
\\ \hline 
3 & 4 & 0.6 & $1.39\times 10^{13}$ & $7.63\times 10^{12}$ & 0.1366 & 0.1302 
\\ \hline\hline 
3 & 5 & 0. & $3\times 10^{16}$ & $3\times 10^{16}$ & 0.1260 & 0.0433 \\  
\hline 
3 & 5 & 0.1 & $2.04\times 10^{16}$ & $1.85\times 10^{16}$ & 0.1265 & 0.0448 
\\ \hline 
3 & 5 & 0.2 & $1.04\times 10^{16}$ & $8.53\times 10^{15}$ & 0.1273 & 0.0477 
\\ \hline 
3 & 5 & 0.3 & $3.31\times 10^{15}$ & $2.45\times 10^{15}$ & 0.1288 & 0.0536 
\\ \hline 
3 & 5 & 0.4 & $4.80\times 10^{14}$ & $3.22\times 10^{14}$ & 0.1315 & 0.0676 
\\ \hline 
3 & 5 & 0.5 & $1.91\times 10^{13}$ & $1.16\times 10^{13}$ & 0.1362 & 0.1203 
\\ \hline\hline 
3 & 6 & 0. & $3\times 10^{16}$ & $3\times 10^{16}$ & 0.1260 & 0.0433 \\  
\hline 
3 & 6 & 0.1 & $1.98\times 10^{16}$ & $1.79\times 10^{16}$ & 0.1265 & 0.0448 
\\ \hline 
3 & 6 & 0.2 & $9.09\times 10^{15}$ & $7.44\times 10^{15}$ & 0.1275 & 0.0483 
\\ \hline 
3 & 6 & 0.3 & $2.09\times 10^{15}$ & $1.55\times 10^{15}$ & 0.1295 & 0.0563 
\\ \hline 
3 & 6 & 0.4 & $1.37\times 10^{14}$ & $9.15\times 10^{13}$ & 0.1333 & 0.0816 
\\ \hline 
3 & 6 & 0.5 & $8.96\times 10^{11}$ & $5.44\times 10^{11}$ & 0.1409 & 0.4619 
\\ \hline 
\end{tabular} 
\end{center} 
\caption{As for Table~3 with $\protect\delta =4,5,6$. For $\ln (\Lambda /%
\protect\mu _{0})$ outside the range presented in the Table, the coupling $%
\protect\alpha _{\Lambda }$ becomes larger than unity. As expected, the 
results for $\Lambda $, $\protect\mu _{0}$ and $\protect\alpha _{3}(M_{z})$ 
are identical to the corresponding cases of Table~2, the only difference 
being a renormalisation of the coupling $\protect\alpha _{\Lambda }$.} 
\end{table} 
 
\section{Conclusions} 
 
We showed that to two-loop order the models of \cite 
{dienes_v1,dienes_v2,dienes_v3} with extra space-time dimensions associated 
with scales less than the unification scale increase the value of the strong 
coupling above the MSSM  value. For a very low value of the 
decompactification scale the prediction is unacceptable. However this result 
is  sensitive to the details of the Kaluza Klein thresholds associated with 
the opening up of the extra dimensions and an acceptable value for the 
strong coupling may be obtained if there are different thresholds for states 
carrying different gauge quantum numbers. For low values of the 
decompactification scale the prediction for the strong coupling is extremely 
sensitive to these Kaluza Klein thresholds. As a result the adjustment\ of 
thresholds needed to bring the strong coupling into agreement with 
experiment must be done very precisely to keep the strong coupling within 
the experimental limits. In this case  the success of the simple unification 
prediction is lost and one needs a detailed model of the Kaluza Klein mass 
spectrum to recover a reliable prediction. For high values of the 
decompactification scale the unification prediction tends to the usual one. 
While there is still a dependence on unification scale thresholds, it is 
much milder so in this case an accurate prediction of the value of the
strong coupling  is possible.

\end{document}